\pdfoutput=1
\documentclass[a4paper,twocolumn,accepted=2026-04-19]{quantumarticle}
\pdfoutput=1

\usepackage{hyperref} % force early load to avoid LaTeX hook-order error
\usepackage{graphicx}
\usepackage{ifthen}
\newcommand{\includefig}[2][]{%
  \begingroup
  \edef\qaFigFile{\detokenize{#2}}%
  \IfFileExists{\qaFigFile}{\includegraphics[#1]{\qaFigFile}}{\fbox{Missing file: \texttt{\detokenize{#2}}}}%
  \endgroup
}

\usepackage{bm}
\usepackage{amssymb}
\usepackage{amsfonts}
\usepackage{color} 
\usepackage{amsmath}
\usepackage{algorithm}
\usepackage{algorithmic}
\usepackage{cases}
\usepackage{amsthm}
\usepackage{fancybox}
\usepackage{enumitem}
\usepackage{ulem} 
 
\begin{document}
\title{Phase Transitions and Noise Robustness of Quantum Graph States}
\author{Tatsuya Numajiri}
\affiliation{Department of Physics, Chuo University, 1-13-27 Kasuga, Bunkyo-ku, Tokyo 112-8551, Japan}
\author{Shion Yamashika}
\affiliation{Department of Physics, Chuo University, 1-13-27 Kasuga, Bunkyo-ku, Tokyo 112-8551, Japan}
\affiliation{Department of Engineering Science, The University of Electro-Communications, Tokyo 182-8585, Japan}
\author{Tomonori Tanizawa}
\affiliation{Department of Physics, Chuo University, 1-13-27 Kasuga, Bunkyo-ku, Tokyo 112-8551, Japan}
\author{Ryosuke Yoshii}
\affiliation{Center for Liberal Arts and Sciences, Sanyo-Onoda City University, 1-1-1 Daigaku-Dori, Sanyo-Onoda, Yamaguchi 756-0884, Japan}
\affiliation{Department of Physics, Chuo University, 1-13-27 Kasuga, Bunkyo-ku, Tokyo 112-8551, Japan}
\affiliation{International Institute for Sustainability with Knotted Chiral Meta Matter (WPI-SKCM2), Hiroshima University, Higashi-Hiroshima, Hiroshima 739-8526, Japan}
\author{Yuki Takeuchi}
\thanks{The current affiliation is Information Technology R\&D Center, Mitsubishi Electric Corporation}
\email{Takeuchi.Yuki@bk.MitsubishiElectric.co.jp}
\affiliation{NTT Communication Science Laboratories, NTT Corporation, 3-1 Morinosato Wakamiya, Atsugi, Kanagawa 243-0198, Japan}
\author{Shunji Tsuchiya}
\email{tshunji001c@g.chuo-u.ac.jp}
\affiliation{Department of Physics, Chuo University, 1-13-27 Kasuga, Bunkyo-ku, Tokyo 112-8551, Japan}

\begin{abstract}
Graph states are entangled states that are essential for quantum information processing. As experimental advances enable the realization of large-scale graph states, efficient fidelity estimation methods are crucial for assessing their robustness against noise. However, calculations of exact fidelity become intractable for large systems due to the exponential growth in the number of stabilizers.
In this work, we show that the fidelity between any ideal graph state and its noisy counterpart under IID Pauli noise can be mapped to the partition function of a classical spin system, enabling efficient computation via statistical mechanical techniques. Using this approach, we analyze the fidelity for regular graph states under depolarizing noise and uncover the emergence of phase transitions in fidelity between the pure-state regime and the noise-dominated regime.
Specifically, in 2D, phase transitions occur only when the degree satisfies $d\ge 6$, while in 3D they already appear at $d\ge 5$. 
However, for graph states with excessively high degree, such as fully connected graphs, the phase transition disappears.
Robustness of graph states against noise is thus determined by their connectivity and spatial dimensionality. Graph states with lower degree and/or dimensionality, which exhibit a smooth crossover, demonstrate greater robustness, while highly connected or higher-dimensional graph states are more fragile. Extreme connectivity, as the fully connected graph state possesses, restores robustness.
Furthermore, we show that the fidelity can be rewritten in the form of the partition function of a constraint-percolation problem. Within this picture, we discuss the qualitative difference between 2D regular graph states with $d= 6$ and $d=5$ regarding the presence or absence of a phase transition, as well as the suppressed critical behavior of fully connected graph states.

\end{abstract}
\maketitle

\section{Introduction}
\label{I}

Graph states are a class of entangled states that serve as essential resources for various quantum information processing tasks, including quantum metrology \cite{SM20}, quantum communication \cite{ATL15}, quantum error correction \cite{SW01}, and measurement-based quantum computation (MBQC) \cite{RB01,RBB03,ZH22,CYZ24}. Given their practical importance, significant efforts have been devoted to generating large-scale graph states, both theoretically \cite{SMSN12,BHKM12,MRRBMDK14,ITTIY14,GSBR15} and experimentally \cite{LZGGZYGYP07,TKYKI08,YWXLPBPLCP12,YWCGFRCLLDCP12,WCLHLCLSWLLHJPLLCLP16,WLHCSLCLFJZLLLP18,WLYZ18,GCZWZDYRWLCZLLXGSCWPLZP19,MHH19, RGYR23}. As experimental capabilities advance to realize increasingly larger graph states, it becomes crucial to develop efficient methods for assessing their fidelity under noise \cite{HM15,MNS16,FH17,PLM18,HH18,TM18,TMMMF19,ZH19,ZH19A,MK20,DHZ20,LZH23,ATYT22,TTYYT23,YS22}.  

While fidelity can be estimated by measuring a polynomial number of randomly sampled stabilizers, avoiding exponential overhead, the accuracy of such estimation protocols must still be rigorously validated against the exact fidelity \cite{ATYT22,TTYYT23}. 
However, computing the exact fidelity requires summing over an exponentially large number of stabilizer expectation values, making the task intractable for large graph states. Thus, developing efficient methods for fidelity evaluation remains a challenging task.

IID Pauli noise, including depolarizing noise, is a widely used theoretical model due to its simplicity
\cite{NC00}. Although it neglects spatial and temporal correlations present in realistic settings, it provides a tractable framework for analyzing the impact of noise on quantum systems. Assessing the reliability of fidelity estimation under such noise requires exact fidelity calculations. 
While this becomes increasingly challenging as the graph state size grows, there are special cases, such as fully connected graph states, where fidelity can be evaluated rather easily \cite{TTYYT23}. In previous studies, mappings of stabilizer states, including 2D \cite{FBAV24} and 3D Toric codes \cite{L24} and the 2D decorated cluster state \cite{GA24}, under IID Pauli errors to classical spin models have been developed. However, these studies are limited to a special type of noise involving independent bit-flip and phase-flip errors.

In this work, we show that the fidelity of any graph state under {\it arbitrary} IID Pauli noise can be mapped to the partition function of a classical spin system, allowing efficient evaluation through statistical mechanical techniques such as transfer matrix methods and Monte Carlo simulations. Using this approach, we compute the fidelity of 1D, 2D, and 3D cluster states, as well as regular graph states, under depolarizing noise and reveal a fundamental connection between graph structures, phase transitions, and noise robustness.

We find that fidelity exhibits a sharp phase transition between the pure-state and noise-dominated regimes, depending on the degree $d$ of the graph state (i.e., the number of edges per qubit) and spatial dimensionality. In 2D, phase transitions appear for $d\ge 6$, while in 3D, they emerge at $d\ge 5$. 
This phase transition arises due to the sudden onset of the Pauli noise contribution at the critical value of probability $p_{\rm c}\simeq 0.5$. For $p\le p_{\rm c}$, the fidelity is dominated by the pure-state contribution $(1-p)^n$.

Furthermore, robustness of graph states against noise is determined by both the connectivity and spatial dimensionality of the graphs. Graph states with lower degree and/or dimensionality exhibit smoother fidelity crossover and are more robust to noise than highly connected and/or high-dimensional ones with phase transitions. Furthermore, higher-dimensional graph states tend to be more fragile under noise, while extremely high connectivity, as in fully connected graphs, suppresses critical behavior and restores robustness.

The paper is organized as follows: Section 2 introduces the formalism for graph states under IID Pauli noise. Section 3 establishes the mapping between the fidelity of graph states and the partition functions of the corresponding classical spin systems. 
Section 4 shows the mean-field results of the corresponding classical spin model. 
In Section 5, we analyze the fidelity of 1D cluster states under depolarizing noise using the transfer matrix method. 
Section 6 extends the study to 2D and 3D cluster states, where we compute fidelity using Monte Carlo simulations. 
Finally, Section 7 summarizes our findings and discusses potential future research directions.

\medskip
\section{Graph states under IID Pauli noise}
\label{II}
In this section, we introduce graph states under IID Pauli noise. We first define the graph states~\cite{RB01}.
A graph $G\equiv(V,E)$ is a pair of the set $V$ of $n$ vertices and the set $E$ of edges that connect the vertices.
The $n$-qubit graph state $|G\rangle$ for the graph $G$ is defined as
\begin{eqnarray}
\label{graph}
|G\rangle\equiv\left(\prod_{(i,j)\in E}CZ_{i,j}\right)|+\rangle^{\otimes n},
\end{eqnarray}
where $|+\rangle\equiv(|0\rangle+|1\rangle)/\sqrt{2}$ with $|0\rangle$ and $|1\rangle$ being, respectively, eigenstates of the Pauli-$Z$ operator with eigenvalues $+1$ and $-1$, and $CZ_{i,j}$ is the controlled-$Z$ ($CZ$) gate applied on the $i$th and $j$th qubits.
The stabilizer generators $\{g_i\}_{i=1}^n$ for $|G\rangle$ are defined as
\begin{eqnarray}
\label{gstabilizer}
g_i\equiv X_i\left(\prod_{j:\ (i,j)\in E}Z_j\right).
\end{eqnarray} 
Here, $X_i$ and $Z_j$ are the Pauli-$X$ and $Z$ operators for the $i$th and $j$th qubits, respectively, and the product of $Z_j$ is taken over all vertices $j$ such that $(i,j)\in E$.
For any $i$ and $j$, two stabilizer generators commute, i.e., $[g_i,g_j]=0$.
The graph state $|G\rangle$ is the unique common eigenstate of $\{g_i\}_{i=1}^n$ with eigenvalue $+1$ , i.e., $g_i|G\rangle=|G\rangle$ for any $i$.

A stabilizer $S_{\ell}$ is a product of stabilizer generators such that $S_{\ell}\equiv\prod_{i=1}^ng_i^{\ell_i}$, where ${\ell}\equiv \ell_1\ell_2\ldots \ell_n\in\{0,1\}^n$.
It is a tensor product of $n$ %single-qubit
Pauli operators with a sign $+$ or $-$. 
More specifically, it can be written as
\begin{equation}
    S_{\ell}=(-1)^s\bigotimes_{i=1}^n\tau_i,\quad \tau_i\in\{I,X,Y,Z\},
    \label{stabilizer}
\end{equation}
where $s\in\{0,1\}$. The two-dimensional identity operator and Pauli-$Y$ operator are denoted as $I$ and $Y=iXZ$, respectively.
For any $\ell$, $S_{\ell}|G\rangle=|G\rangle$ holds due to Eqs.~(\ref{graph}) and (\ref{gstabilizer}).

The IID Pauli noise is represented by the superoperator \cite{NC00}
\begin{eqnarray}
\mathcal{E}(\cdot)\equiv&(1-p)I(\cdot)I+p_xX(\cdot)X
+p_yY(\cdot)Y\notag\\&+p_zZ(\cdot)Z,
%\label{dplns}
\label{lPns}
\end{eqnarray}
where $p\equiv p_x+p_y+p_z$. It operates independently on each qubit, where bit-flip (Pauli-$X$ error), phase-flip (Pauli-$Z$ error), and bit-phase-flip errors (Pauli-$Y$ error) occur with probabilities $p_x$, $p_z$, and $p_y$, respectively.

The IID Pauli noise includes the depolarizing noise as a special case when $p_x=p_y=p_z=p/3$.
The depolarizing noise interpolates between the original state and the maximally mixed state \cite{NC00} as
\begin{align}
\mathcal{E}(\rho)=(1-p')\rho+p'\frac{I}{2},
\label{eq:deponoise}
\end{align}
where $0\le p'\le 1$ and $p=\frac{3}{4}p'$. 
Given the expression Eq.~(\ref{eq:deponoise}), we restrict $p$ within the region $0\le p\le 3/4$ in the following.

Let $\ovalbox{$|\psi\rangle$}\equiv|\psi\rangle\langle\psi|$ for any pure state $|\psi\rangle$.
The density operator $\rho\equiv\mathcal{E}^{\otimes n}(|G\rangle\langle G|)$ for the graph state under IID Pauli noise is written as 
\begin{align}
&\rho=(1-p)^n\ovalbox{$|G\rangle$}
+\sum_{m=1}^n(1-p)^{n-m}\notag\\
&\times\sum_{\substack{1\le i_1<i_2<\ldots<i_m\le n \\ \mu_1,\mu_2,\ldots,\mu_m}}
%\left(\prod_{k=1}^m p_{\mu_k}\right)
p_x^{m_x}p_y^{m_y}p_z^{m_z}
\ovalbox{$(\prod_{l=1}^m\sigma_{\mu_l i_l})|G\rangle$}~,
\label{dperror}
\end{align}
where $\sigma_{1i}\equiv X_i$, $\sigma_{2i}\equiv Y_i$, and $\sigma_{3i}\equiv Z_i$, and the summation of $\mu_i$ ($1\le i\le m$) is taken over $\mu_i=1,2,3$. We denote the numbers of $X$, $Y$, and $Z$ in the product $\prod_{l=1}^m\sigma_{\mu_l i_l}$ as $m_x$, $m_y$, and $m_z$, respectively, then $m_x+m_y+m_z=m$.

The fidelity $\langle G|\rho|G\rangle$ between the graph state under the IID Pauli noise $\rho$ and the ideal state $|G\rangle\langle G|$ can be written as
\begin{align}
F&=(1-p)^n+
\sum_{m=1}^n(1-p)^{n-m}\notag\\
&\times \sum_{\substack{1\le i_1<\ldots <i_m\le n \\ \mu_1,\ldots,\mu_m}}p_x^{m_x}p_y^{m_y}p_z^{m_z}\langle G|\left(\prod_{k=1}^m\sigma_{\mu_ki_k}\right)|G\rangle^2,
\label{eq:fidelity}
\end{align} 
where the summation is taken only over $i_1$ and $\mu_1$ in the case of $m=1$. 
The expression
\begin{eqnarray}
\sum_{\substack{1\le i_1<\ldots<i_m\le n \\ \mu_1,\ldots,\mu_m}}\langle G|\left(\prod_{k=1}^m\sigma_{\mu_ki_k}\right)|G\rangle^2
\end{eqnarray}
corresponds to the number of stabilizers that contain exactly $m_x$, $m_y$, and $m_z$ operators for $X$, $Y$, and $Z$, respectively. This is because, for the expectation value $\langle G|\left(\prod_{k=1}^m\sigma_{\mu_ki_k}\right)|G\rangle$ to be nonzero, the operator $\prod_{k=1}^m\sigma_{\mu_k i_k}$ must match a stabilizer (with either a $+$ or $-$ sign) from Lemma 1 in Ref.~\cite{TTYYT23}. 
We note that the second term on the right-hand side of Eq.~(\ref{eq:fidelity}) is always positive, indicating that the fidelity increases when the product of Pauli noise operators, $\prod_{k=1}^m\sigma_{\mu_k i_k}$, coincides with a stabilizer of the graph state.

In general, the number of stabilizers increases exponentially with $n$. As a result, determining the exact value of $F$ in Eq.~(\ref{eq:fidelity}) becomes increasingly difficult for large $n$, since it requires counting the number of stabilizers containing specified numbers of Pauli $X$, $Y$, and $Z$ operators, i.e., for given values of $m_x$, $m_y$, and $m_z$. However, efficient computation is still possible by leveraging the mapping of the fidelity to the partition function of an equivalent classical spin system and using methods of statistical mechanics, as described in later sections.

\medskip
\section{Mapping Fidelity of a graph state to the Partition Function of a Classical Spin System}
\label{III}

In this section, we demonstrate that the fidelity $F$ between a graph state under the IID Pauli noise $\rho$ and an ideal graph state $|G\rangle\langle G|$ can be mapped to the partition function of a classical spin system. Our starting point is the expression of the fidelity in Eq.~(\ref{eq:fidelity}):
\begin{equation}
F = \sum_{m=0}^{n} (1 - p)^{n - m} p_x^{m_x}p_y^{m_y}p_z^{m_z} N_F(m_x,m_y,m_z).
\label{eq:fidelity_short}
\end{equation}
$N_F(m_x,m_y,m_z)$ denotes the number of stabilizers that are products of $m_x$, $m_y$, and $m_z$ operators ($m_x,m_y,m_z\in \mathbb{N}_0$) for $X$, $Y$, and $Z$, respectively. 

The key idea behind the mapping is as follows: The fidelity is evaluated by summing the number of stabilizers, weighted by their probabilities, over the number of Pauli operators $m_x$, $m_y$, and $m_z$ in Eq.~(\ref{eq:fidelity_short}). This is analogous to computing the partition function of a classical spin system, such as the Ising model, by summing the number of spin configurations, weighted by their Boltzmann factors, over the total energy of the spin configurations. In this mapping, each stabilizer corresponds to a spin configuration, the number of Pauli $X$, $Y$, and $Z$ operators in a stabilizer corresponds to the energy of the corresponding spin configuration, and the probability corresponds to the Boltzmann factor.

Based on this idea, we follow the four steps below to map the expression Eq.~(\ref{eq:fidelity_short}) to the partition function of a classical spin system.

\begin{figure*}[t]
\begin{center}
\includefig[width=14cm,clip]{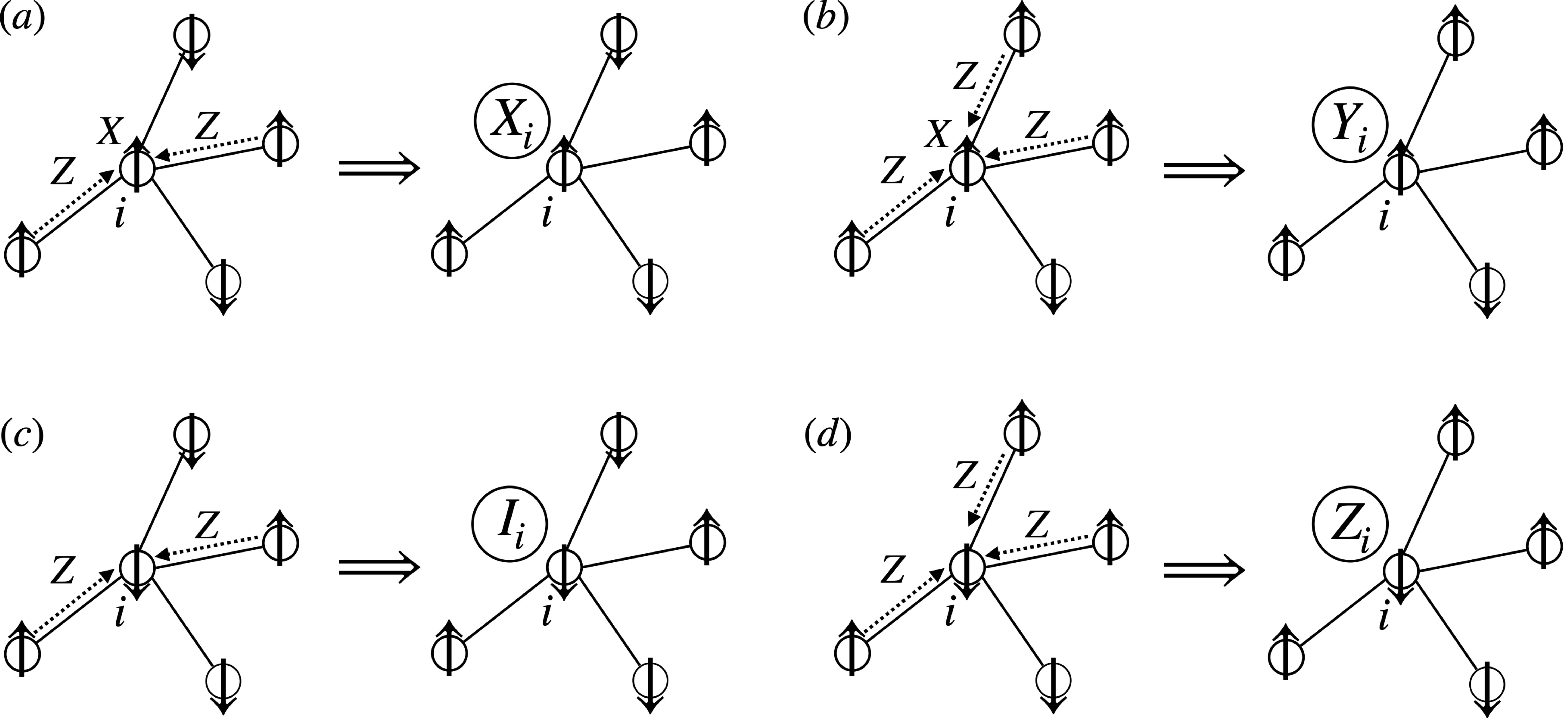}
\caption{Illustration of how the local operator $\tau_i\in\{X,Y,Z,I\}$ acting on qubit $i$ in a stabilizer $S_\ell$ is determined by the presence or absence of $g_i$ ($s_i=\pm 1$) and the parity of the number of stabilizer generators (spin $\uparrow$s) connected to qubit $i$. In the presence of $g_i$, (a) $\tau_i=X$ when the number of generators is even, and (b) $\tau_i=Y$ when the number of generators is odd. In the absence of $g_i$, (c) $\tau_i=I$ when the number of generators is even, and (d) $\tau_i=Z$ when the number of generators is odd. }
\label{fig:rule}
\end{center}
\end{figure*}

\begin{enumerate}[leftmargin=0.3cm]
\item {\bf Introduction of Classical Spins}\\
For a stabilizer $S_{\ell}=\prod_{i=1}^n g_i^{\ell_i}$, we define a classical spin corresponding to the qubit $i$ as $s_i=2\ell_i-1$. 
That is, if $S_{\ell}$ includes $g_i$ (where $\ell_i=1$), then $s_i=+1$; if $S_{\ell}$ does not include $g_i$ (where $\ell_i=0$), then $s_i=-1$. Thus, for a given $S_{\ell}$, the corresponding spin configuration is determined.

\item {\bf Construction of the Hamiltonian}\\
For the stabilizer $S_{\ell}$ in Eq.~ (\ref{stabilizer}), we determine under which conditions the Pauli operator $\tau_i$ becomes $X$, $Y$, $Z$, or $I$.
\begin{enumerate}[leftmargin=0.5cm]
\item[(i).] Cases where $S_{\ell}$ includes $g_i$ ($s_i=+1$):\\
In this case, $\tau_i$ is either $X$ or $Y$. If $S_{\ell}$ has an even number of generators (up spins) on the qubits connected to qubit $i$ by edges, then $\tau_i=X_i$, because the even number of $Z_i$s from these generators cancel out, leaving only $X_i$ from $g_i$ (Fig.~\ref{fig:rule} (a)). Conversely, if $S_{\ell}$ has an odd number of such generators, then $\tau_i=Y_i$, since both $Z_i$ from such generators and $X_i$ from $g_i$ remain (Fig.~\ref{fig:rule} (b)). Thus, the numbers of $X$ and $Y$ in $S_{\ell}$, denoted by ${\mathcal H}_X$ and ${\mathcal H}_Y$ respectively, are given by
\begin{flalign}
&{\mathcal H}_X = \sum_{i=1}^{n} \frac{1}{2}(1 + s_i) \frac{1}{2}\big(1 + \prod_{j: (i,j)\in E} (-s_j)\big),\label{numberofX}\\
&{\mathcal H}_Y = \sum_{i=1}^{n} \frac{1}{2}(1 + s_i) \frac{1}{2}\big(1 - \prod_{j: (i,j)\in E} (-s_j)\big).\label{numberofY}
\end{flalign}
Here, $(1+s_i)/2$ and $(1 \pm \prod_{j: (i,j)\in E} (-s_j))/2$ act as projection operators onto the subspaces with $s_i=+1$ and with an even (odd) number of up spins connected to qubit $i$ by edges, respectively.
Thus, ${\mathcal H}_X$ and ${\mathcal H}_Y$ have eigenvalues $m_x$ and $m_y$, respectively.
\item[(ii).] Cases where $S_{\ell}$ does not include $g_i$ ($s_i=-1$):\\
In this case, $\tau_i$ is either $I$ or $Z$. If $S_{\ell}$ has an even number of generators on the qubits connected to qubit $i$ by edges, then $\tau_i=I_i$ (Fig.\ref{fig:rule} (c)); if $S_{\ell}$ has an odd number of such generators, then $\tau_i=Z_i$ (Fig.\ref{fig:rule} (d)). Thus, the numbers of $Z$ and $I$ in $S_{\ell}$, denoted by ${\mathcal H}_Z$ and ${\mathcal H}_I$ respectively, are given by
\begin{flalign}
{\mathcal H}_Z = \sum_{i=1}^{n} \frac{1}{2}(1 - s_i) \frac{1}{2}\big(1 - \prod_{j: (i,j)\in E} (-s_j)\big),\label{numberofZ}\\
{\mathcal H}_I = \sum_{i=1}^{n} \frac{1}{2}(1 - s_i) \frac{1}{2}\big(1 + \prod_{j: (i,j)\in E} (-s_j)\big).\label{numberofI}
\end{flalign}
Here, ${\mathcal H}_Z$ and ${\mathcal H}_I$ have eigenvalues $m_z$ and $m_i=n-(m_x+m_y+m_z)$, respectively.
\end{enumerate}

Based on (i) and (ii), the Hamiltonian with energy eigenvalues that depend on the numbers of $X$, $Y$, and $Z$ in $S_{\ell}$ can be written as 
\begin{equation}
\mathcal H = J_x{\mathcal H}_X+J_y{\mathcal H}_Y+J_z{\mathcal H}_Z,
\label{Hamiltonian}
\end{equation}
where the ratio between $J_\mu$ ($\mu=x,y,z$) are determined by $p_x$, $p_y$, and $p_z$, as we discuss in the next step. Its energy eigenvalues are denoted as $E(m_x,m_y,m_z)\equiv J_x m_x+J_ym_y+J_zm_z$. The number of spin configurations that have an energy eigenvalue $E(m_x,m_y,m_z)$ is equal to $N_F(m_x,m_y,m_z)$. The ground state of the Hamiltonian with $J_\mu>0$ ($\mu=x,y,z$), regardless of the shape of the graph state, is the ferromagnetic state where all spins are $-1$, and the energy eigenvalue is zero, corresponding to $S_{\ell}=I^{\otimes n}$.

\item {\bf Determination of Parameters}\\
The partition function $\mathcal Z$ for the classical spin system $\{s_i\}_{i=1}^n$ that is described by the Hamiltonian (\ref{Hamiltonian}) can be written as
\begin{equation}
\mathcal Z = \sum_{m=0}^{n} e^{-\beta({E(m_x, m_y, m_z)}+c)}N_F(m_x,m_y,m_z).
\label{eq:partition_function}
\end{equation}
Here, $\beta$ is the inverse temperature and $c$ is an energy offset. 
The free energy of the classical spin system is given as
\begin{equation}
    \mathcal F=-\frac{1}{\beta}\ln\mathcal Z.
\end{equation}

We want to map this $\mathcal Z$ to the fidelity expression in Eq.~(\ref{eq:fidelity_short}).
For this, since the Boltzmann factor in Eq.~(\ref{eq:partition_function}) corresponds to the probability in Eq.~(\ref{eq:fidelity_short}), it should hold that
\begin{align}
&\beta J_\mu=\ln\left(\frac{1-p}{p_\mu}\right),\quad (\mu=x,y,z),
\label{eq:J(p)}\\
&c=-\frac{n}{\beta}\ln(1-p).
\end{align}
\end{enumerate}
Note that Eq.~(\ref{eq:J(p)}) can determine the ratio between $J_\mu$ ($\mu=x,y,z$). Fixing one of them by hand, say $J_x=1$, $\beta$ can be uniquely determined.

For depolarizing noise, where $p_x = p_y = p_z = p/3$, setting $J_x = J_y = J_z = 1$, the Hamiltonian is expressed as
\begin{align}
&\mathcal H={\mathcal H}_X+{\mathcal H}_Y+{\mathcal H}_Z\notag\\
&=\sum_{i=1}^n\left[\frac{1}{2}(1+s_i)+\frac{1}{4}(1-s_i)\{1-\prod_{j: (i,j)\in E}(-s_j)\}\right].
\label{eq:depolarizingH}
\end{align}
This Hamiltonian has eigenvalues that are equal to the total number of Pauli $X$, $Y$, and $Z$ operators in $S_\ell$. The inverse temperature $\beta$ of the equivalent classical spin system is given by
\begin{eqnarray}
\beta=\ln{\frac{3(1-p)}{p}}.
\label{eq:beta}
\end{eqnarray}  

\begin{figure}[t]
\includefig[width=6cm,clip]{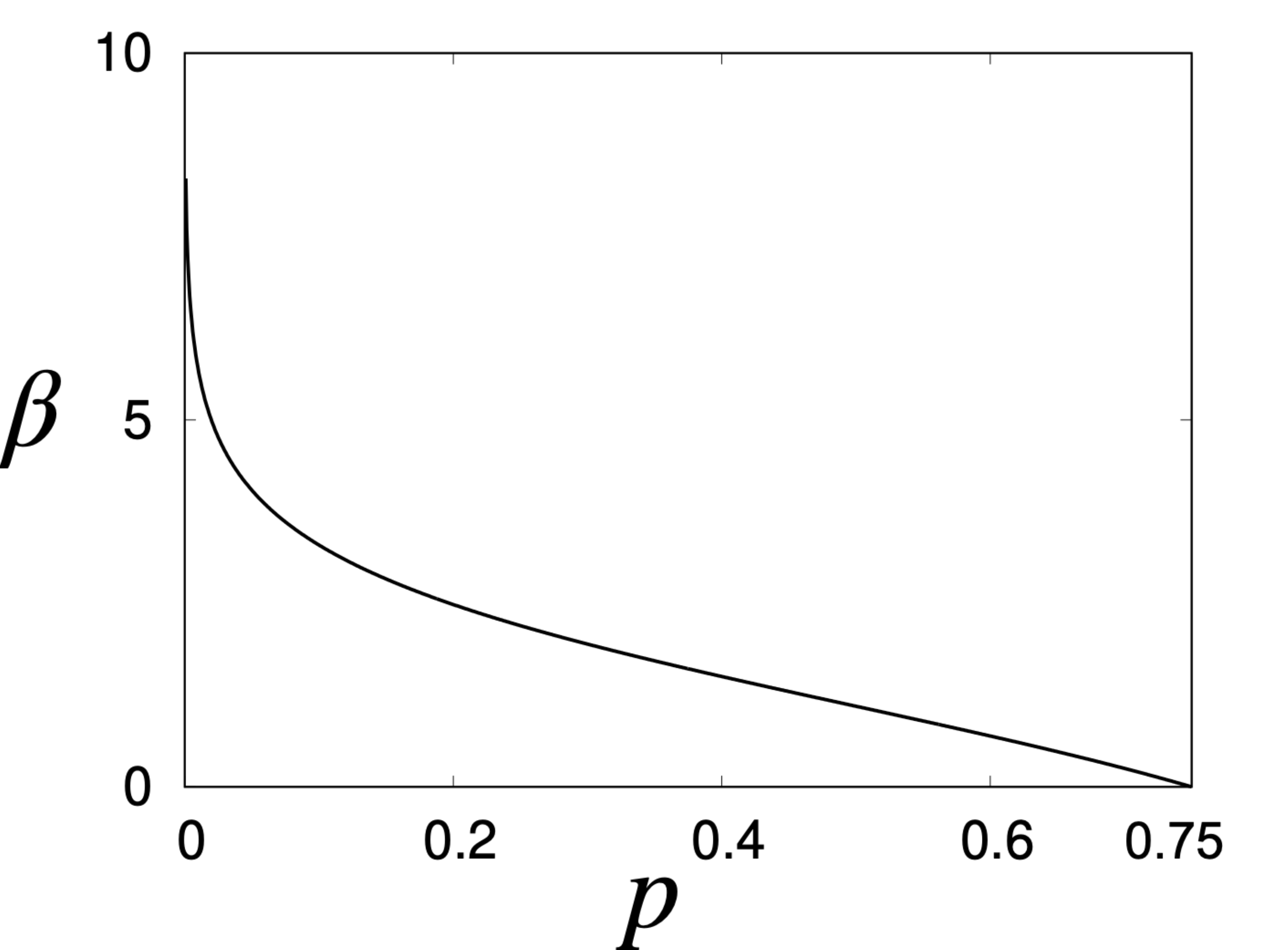}
\caption{The $p$-dependence of the inverse temperature $\beta$: $p\to 0$ corresponds to the low-temperature region ($\beta\rightarrow\infty$), while $p\to 3/4$ corresponds to the high-temperature region ($\beta\rightarrow 0$).}
\label{fig:beta}
\end{figure}

Figure~\ref{fig:beta} plots $\beta$ in Eq.~(\ref{eq:beta}) as a function of the error probability $p$. $p=0$ corresponds to the zero temperature limit ($\beta \to \infty$). For $0<p<3/4$, $\beta$ decreases as $p$ increases. 
At $p = 3/4$, the system reaches the infinite-temperature limit ($\beta = 0$), where the density matrix $\rho$ becomes the maximally mixed state $\rho=I^{\otimes n}/2^n$ from Eq.~(\ref{eq:deponoise}). 

\section{Mean-field results}

In this and the following sections, we focus on graph states subject to depolarizing noise. We note, however, that the same analytical and numerical frameworks can be applied to evaluate the fidelity under other types of IID Pauli noise as well.

We first evaluate the fidelity of 1D, 2D, and 3D cluster states using the mean-field (MF) approximation to the classical spin Hamiltonian for depolarizing noise (\ref{eq:depolarizingH}). We describe the details of the MF theory in Appendix A.  

\begin{figure*}[t]
\includefig[width=\textwidth,clip]{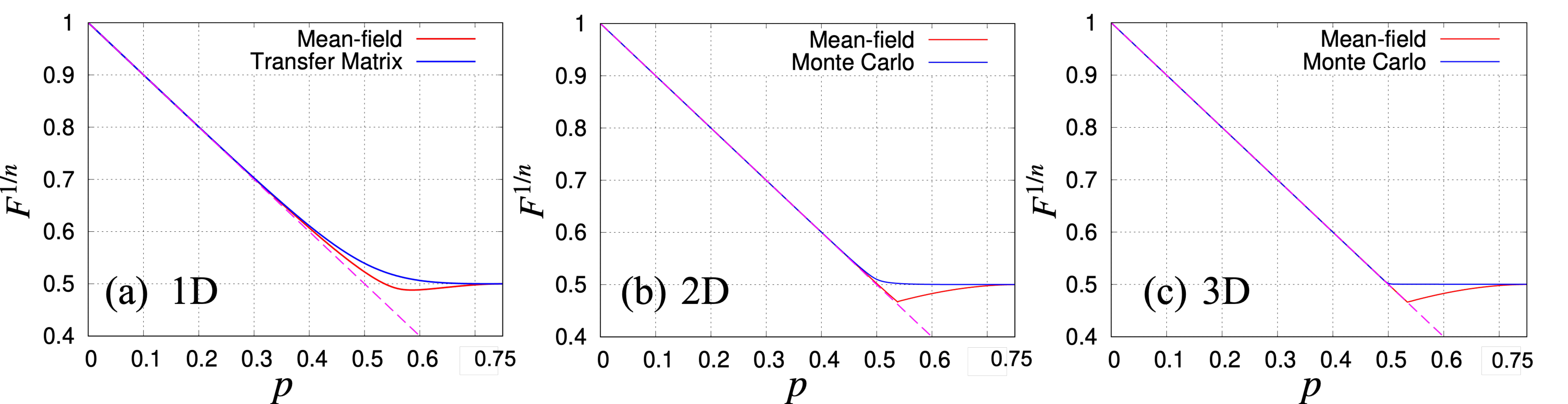}
\caption{Mean-field results for the fidelity $F^{1/n}$ of 1D (a), 2D (b), and 3D (c) cluster states under depolarizing noise (red curves). The blue curves show the corresponding results obtained using the transfer matrix method for the 1D cluster state with $n=1000$ qubits (a), and Monte Carlo simulations for the 2D cluster state with $n_x=n_y=60$ (b), and the 3D cluster state with $n_x=n_y=n_z=6$ (c). The dashed magenta lines correspond to $1-p$.}
\label{fig:meanfield}
\end{figure*} 

The MF results of the fidelity in Figs.~\ref{fig:meanfield} (a), (b), and (c) indicate that $F^{1/n}$ exhibits a singularity point at $p_{\rm c}^{\rm MF} = 0.535$ and $0.538$ for the 3D and 2D cluster states, respectively, while no such point is observed for the 1D cluster state. These singularities suggest that a phase transition occurs at $p_{\rm c}^{\rm MF}$. For $p < p_{\rm c}^{\rm MF}$, $F^{1/n}$ decreases linearly with $1-p$. 

Since the MF free energy is 
%larger than 
larger than or equal to 
the actual free energy $\mathcal F_{\rm MF}\ge \mathcal F$ due to the Gibbs-Bogoliubov-Feynmann inequality \cite{B58, F55}, the mean-field partition function provides a lower bound for the actual partition function, that is, $\mathcal{Z} \ge \mathcal{Z}_{\rm MF}$. 
Consequently, fidelity within the MF approximation also provides a lower bound for actual fidelity, i.e., $F \ge F_{\rm MF}$, as shown in Figs.~\ref{fig:meanfield} (a), (b), and (c).

\section{Transfer matrix approach for 1D cluster state}

\begin{figure*}[t]
\includefig[width=\linewidth,clip]{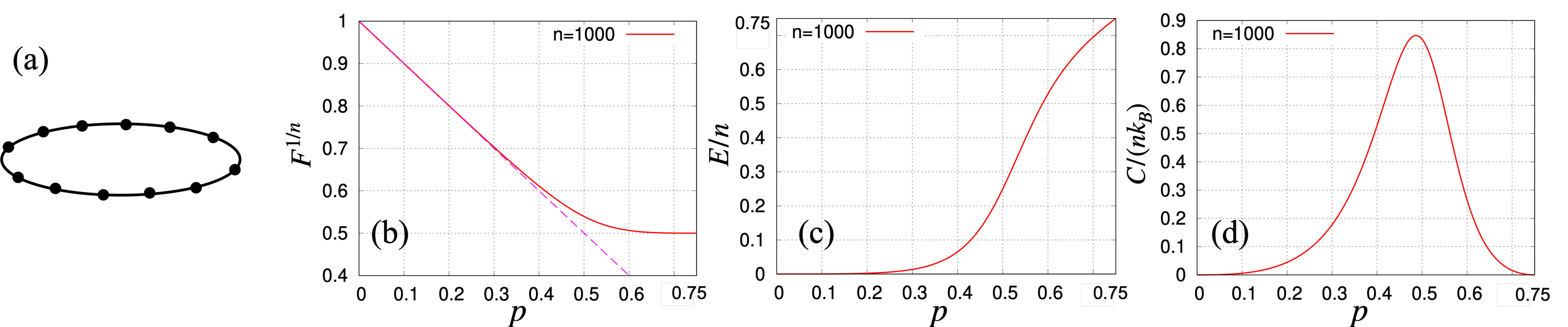}
\caption{(a) 1D cluster state, where black dots represent qubits and black lines indicate $CZ$ gates. (b) Fidelity, (c) internal energy, and (d) specific heat as functions of the noise parameter 
$p$ for the 1D cluster state with number of qubits $n=1000$. The solid red curves in (b), (c), and (d) show the results obtained using the transfer matrix method. The dashed magenta line in (b) corresponds to $1-p$. $k_{\rm B}$ denotes the Boltzmann constant.}
\label{fig:1Dcluster}
\end{figure*}

The standard transfer matrix approach is applicable to the 1D cluster state under the periodic boundary condition shown in Fig.~\ref{fig:1Dcluster} (a). The Hamiltonian can be written as
\begin{align}
&\mathcal H=\sum_{i=1}^{n} h_{i}=\sum_{i=1}^{n/2}\tilde h_{2i-1,2i},
\end{align}
where we denote
\begin{align}
&h_{i}
=\frac{1}{2}(1+s_i)+\frac{1}{4}(1-s_i)(1-s_{i-1}s_{i+1}),\\
&\tilde h_{2i-1,2i}=h_{2i-1}+h_{2i}.
\end{align}
Note that $h_i=h_i(s_{i-1},s_i,s_{i+1})$ and $\tilde h_{2i-1,2i}=\tilde h_{2i-1,2i}(s_{2i-2},s_{2i-1},s_{2i},s_{2i+1})$.
The partition function can be written as
\begin{align}
&\mathcal Z=\sum_{s_1=\pm 1}\dots\sum_{s_n=\pm 1}\exp\left[-\beta\left(\sum_{i=1}^{n/2}\tilde h_{2i-1,2i}+c\right)\right]\notag\\
&=e^{-\beta c}\left[\prod_{j=1}^n\sum_{s_j=\pm 1}\right]\prod_{i=1}^{n/2}T((s_{2i-2},s_{2i-1}),(s_{2i},s_{2i+1})),
\end{align}
where the transfer matrix $T$ is given as
\begin{align}
&T((s_{2i-2},s_{2i-1}),(s_{2i},s_{2i+1}))\notag\\
&=\exp\left(-\beta \tilde h_{2i-1,2i}(s_{2i-2},s_{2i-1},s_{2i},s_{2i+1})\right).
\end{align}
Setting the order of the basis as $(1,1)$, $(1,-1)$, $(-1,1)$, and $(-1,-1)$, the explicit form of $T$ is
\begin{equation}
T=
\begin{pmatrix}
e^{-2\beta} & e^{-2\beta} & e^{-\beta} & e^{-2\beta} \\
e^{-\beta} & e^{-\beta} & e^{-2\beta} & e^{-\beta} \\
e^{-2\beta} & e^{-2\beta} & e^{-\beta} & e^{-2\beta} \\
e^{-2\beta} & e^{-2\beta} & e^{-\beta} & 1
\end{pmatrix}
.
\label{eq:T-matrix}
\end{equation}
Note that one of the eigenvalues of the transfer matrix $T$ is zero, as the third row is identical to the first in Eq.~(\ref{eq:T-matrix}). 
The remaining three eigenvalues are found to be real. 

The partition function is obtained as
\begin{align}
&\mathcal Z=e^{-\beta c}~{\rm Tr}(T^{n/2})\nonumber\\
&=e^{-\beta c}(\lambda_1^{n/2}+\lambda_2^{n/2}+\lambda_3^{n/2}),
\end{align}
where $\lambda_k$ ($k=1,2,3$) are the nonzero eigenvalues of $T$.  
Thus, we obtain the fidelity as
\begin{align}
F^{1/n}=(1-p)(\lambda_1^{n/2}+\lambda_2^{n/2}+\lambda_3^{n/2})^{1/n}.
\end{align}

In addition to fidelity, we also evaluate the internal energy $E=-\frac{d}{d\beta}\ln \mathcal Z$ and the specific heat $C=-\beta^2\frac{dE}{d\beta}$ of the classical spin system as functions of the noise parameter $p$. 
The internal energy, which corresponds to the expectation value of the Hamiltonian Eq.~(\ref{eq:depolarizingH}), reflects the averaged total number of Pauli $X$, $Y$, and $Z$ operators appearing in a stabilizer, since the mapped classical spin Hamiltonian assigns an energy equal to the number of such operators in each stabilizer.

Figures~\ref{fig:1Dcluster} (b)–(d) show the results for the 1D cluster state obtained using the transfer matrix approach. As seen in Fig.~\ref{fig:1Dcluster} (b), the fidelity exhibits a smooth crossover from the pure state to the maximally mixed state around $p= 0.5$. 
Correspondingly, the internal energy increases monotonically with $p$, and the specific heat shows a broad, non-divergent peak near $p=0.5$. These results indicate that no phase transition occurs in the 1D cluster state.

A phase transition is marked by non-analytic behavior of the free energy in the limit $n\to\infty$. In this limit, the partition function and the fidelity become $\mathcal Z\to e^{-\beta c}\lambda_1^{n/2}$ and $F^{1/n}\to (1-p)\lambda_1^{1/2}$, respectively, where $\lambda_1$ is the largest eigenvalue. The free energy, then, becomes
\begin{align}
\mathcal F\to c-\frac{n}{2\beta}\ln\lambda_1.
\end{align}
Since $T$ has strictly positive matrix elements, $\lambda_1$ is positive and non-degenerate from the Perron-Frobenius theorem and hence $\mathcal F$ is analytic. Thus, we can conclude that no phase transition occurs in the fidelity.

The fidelity $F^{1/n}$ in Fig.~\ref{fig:1Dcluster} (b) deviates from and exceeds $1-p$ for $p\gtrsim 0.4$. This deviation arises from the second term in Eq.~(\ref{eq:fidelity}), which accounts for contributions from Pauli noise operators that coincide with stabilizers. Such contributions lead to the smooth crossover between the pure-state regime and the maximally mixed-state regime.

\section{Monte Carlo approach for 2D and 3D cluster states}

For 2D and 3D graph states, the fidelity can be computed by using the Monte Carlo method for the classical spin system. Our Monte Carlo calculation is based on the Metropolis algorithm. 

\begin{figure*}[t]
\includefig[width=12cm,clip]{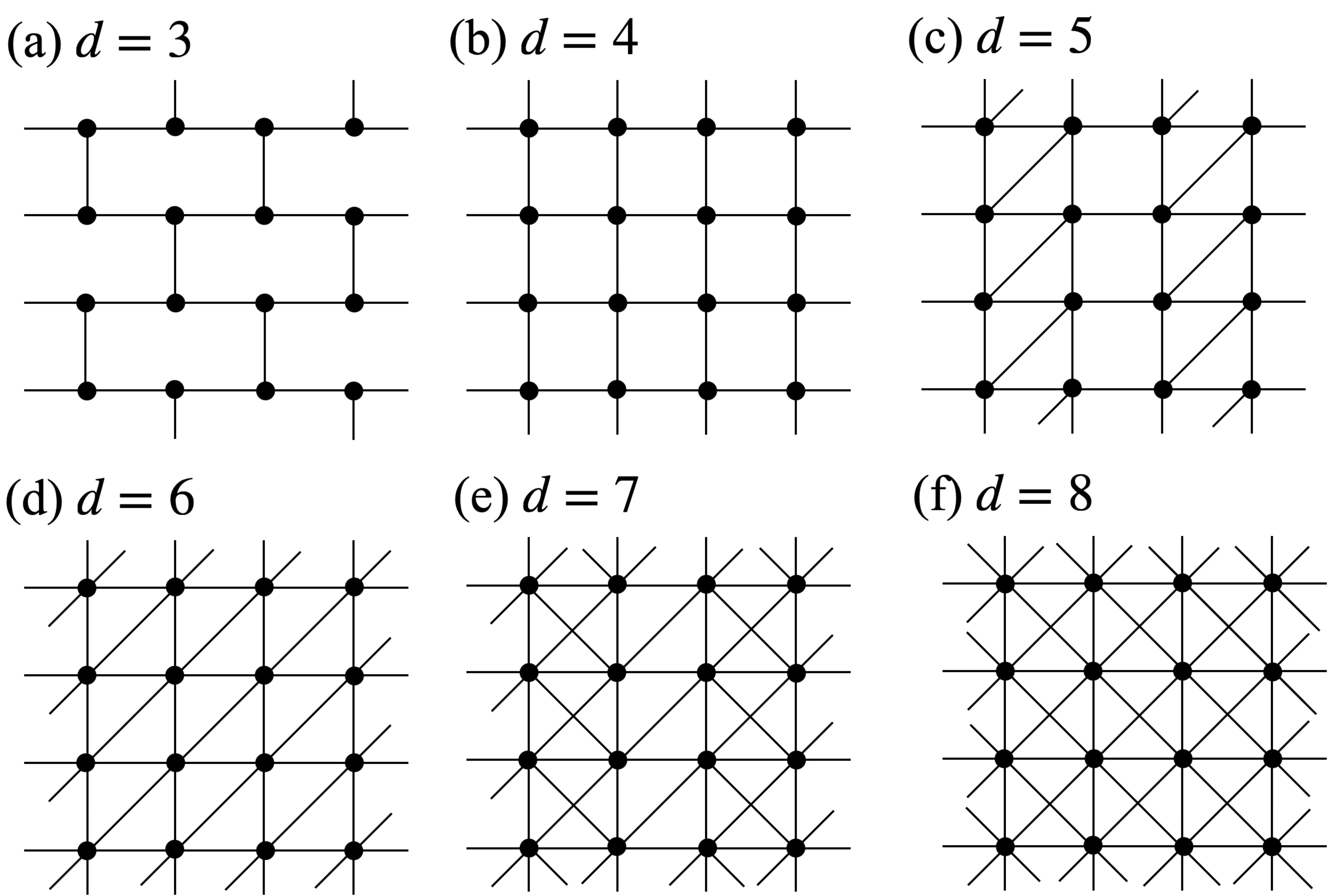}
\caption{2D $d$-regular graph states for $3\le d\le 8$. Each black dot and line indicate a qubit and $CZ$ gate, respectively.}
\label{fig:2Dgraph}
\end{figure*}

We focus on graph states on uniform 2D and 3D $d$-regular graphs, where each vertex is connected to $d$ edges. A $d$-regular graph state is said to have degree $d$. Figures~\ref{fig:2Dgraph} (a)-(f) show 2D $d$-regular graph states for $3\leq d\leq 8$. The 4-regular graph state corresponds to the 2D cluster state. We note that the number of qubits in the transverse direction must be even when $d$ is odd under the periodic boundary condition.

A 3D $(d+2)$-regular graph state can be constructed by stacking 2D $d$-regular graph states and then connecting each qubit to its counterparts in the adjacent upper and lower layers via additional edges. In this way, the 3D cluster state (with degree 6) is obtained by stacking 2D cluster states (degree 4), while the 3D 5-regular graph state is constructed by stacking 2D 3-regular graphs.

\subsection{2D graph states}

\begin{figure*}[t]
\includefig[width=\linewidth,clip]{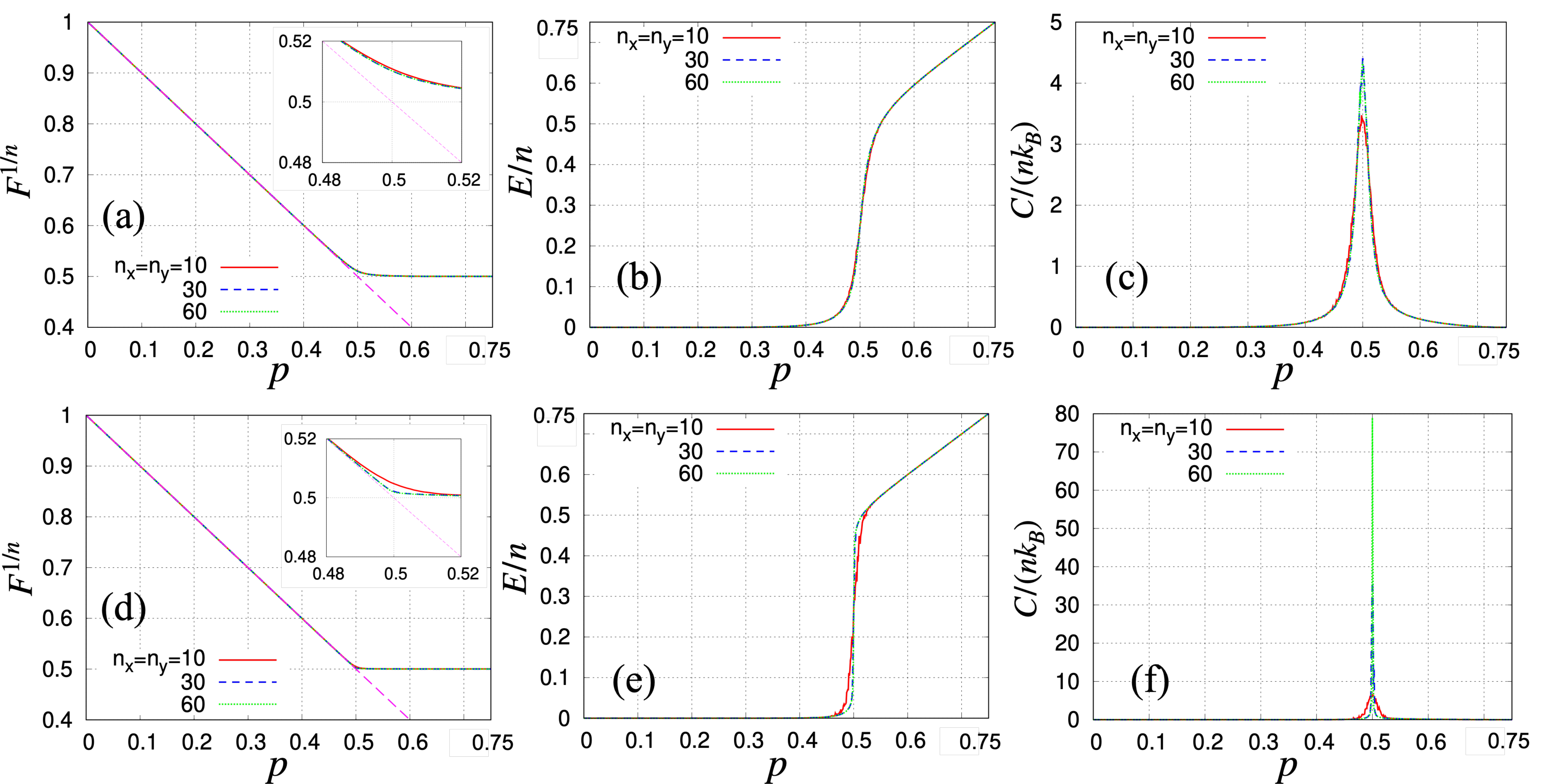}
\caption{Fidelity ((a), (d)), internal energy ((b), (e)), and specific heat ((c), (f)) as functions of the noise parameter $p$, for the 2D cluster state ((a)–(c)) and the 2D 6-regular graph state ((d)–(f)). $n_x$ and $n_y$ denote the numbers of qubits of the 2D graph states in the $x$ and $y$ directions. The insets in (a) and (d) show magnified views of the corresponding plots. The dashed magenta lines in (a) and (d) correspond to $1-p$.}
\label{fig:2Dgraphstate}
\end{figure*}

The results of the 2D cluster state are presented in Figs.~\ref{fig:2Dgraphstate} (a)-(c). As shown in Fig.~\ref{fig:2Dgraphstate} (a), $F^{1/n}$ for the 2D cluster state decreases linearly with $p$, similar to the 1D case, and continuously approaches the value of the maximally mixed state, $1/2$, at $p \simeq 0.5$. 
Although the transition from the linear regime to the constant regime is more abrupt than in the 1D case in Fig.~\ref{fig:1Dcluster} (b), it remains smooth even as the system size increases, indicating the absence of a phase transition. Consistent with this behavior, the internal energy increases smoothly with $p$, and the specific heat exhibits a relatively broad peak near $p=0.5$ that does not grow with the system size (see Figs.~\ref{fig:2Dgraphstate} (b) and (c)).

\begin{figure*}[t]
\includefig[width=13cm,clip]{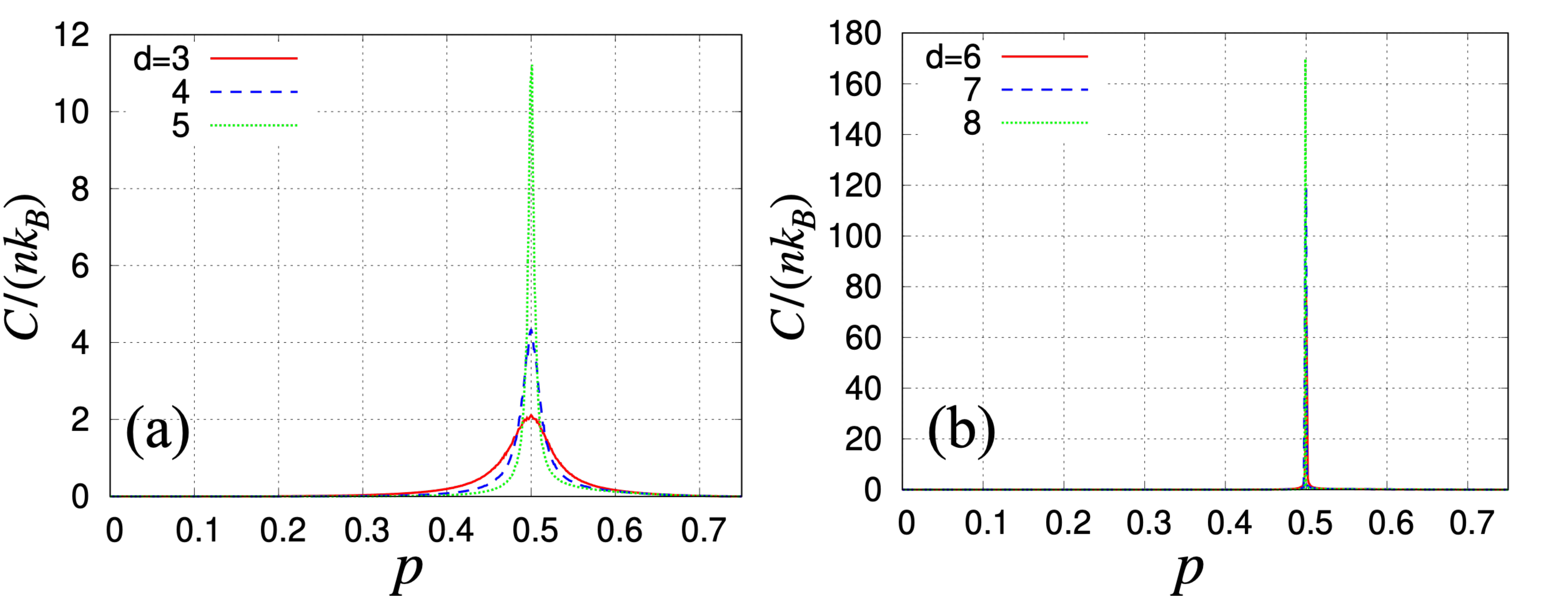}
\caption{Specific heat of 2D uniform graph states with degree $d=3,4,5$ in (a) and $d=6,7,8$ in (b), plotted as functions of the noise parameter $p$, where we set the size of the graph states $n_x=n_y=60$.}
\label{fig:2D_spheat}
\end{figure*}

\begin{figure*}[t]
\includefig[width=\linewidth,clip]{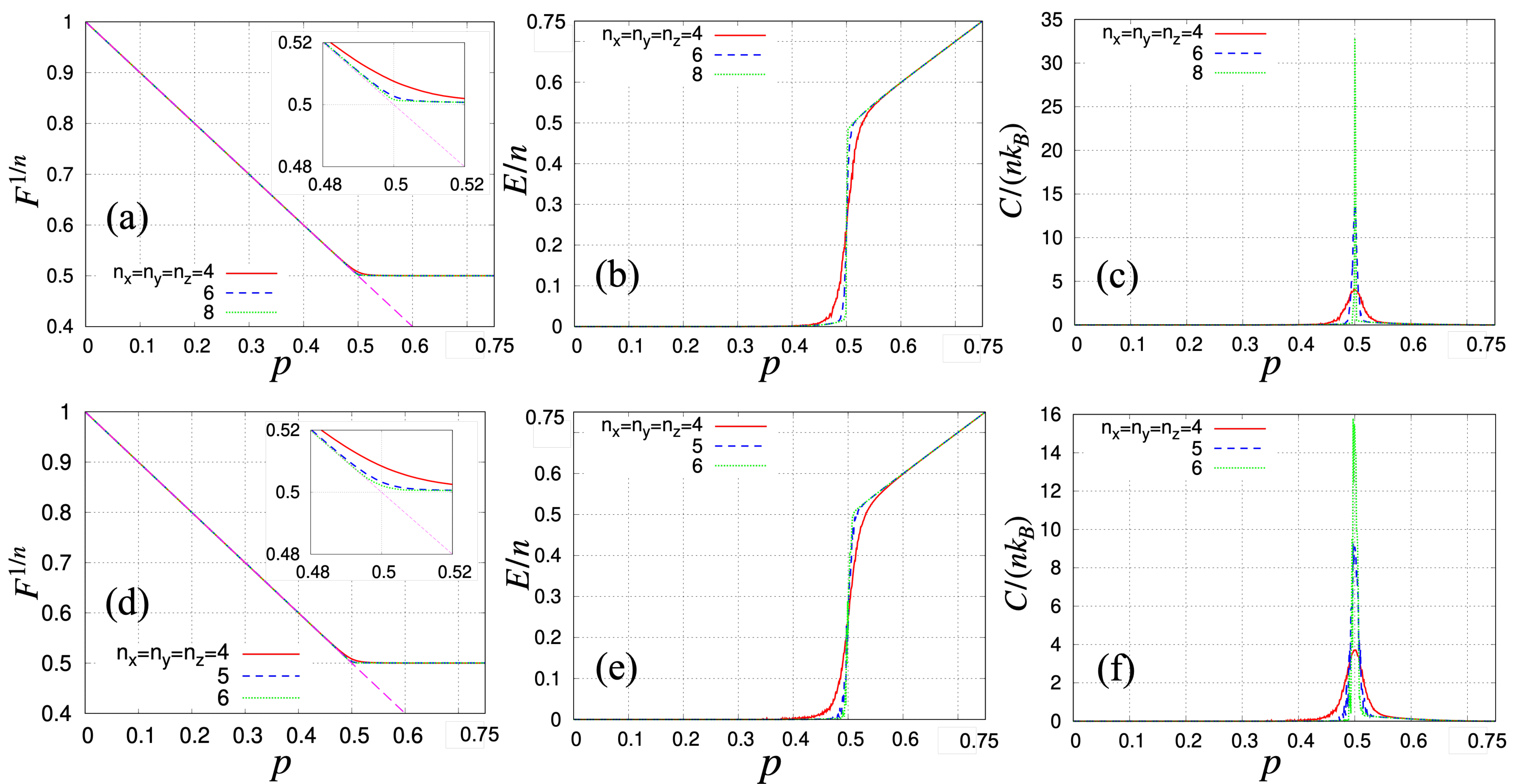}
\caption{Fidelity ((a), (d)), internal energy ((b), (e)), and specific heat ((c), (f)) as functions of the noise parameter $p$, for the 3D 5-regular graph state ((a)–(c)) and the 3D cluster state ((d)–(f)). The insets in (a) and (d) show magnified views of the corresponding plots. The dashed magenta lines in (a) and (d) correspond to $1-p$.}
\label{fig:3Dgraphstate}
\end{figure*}

To compare with the results for the 2D cluster state, Figs.~\ref{fig:2Dgraphstate} (d)–(f) show the corresponding results for the 6-regular graph state. In contrast to the 2D cluster state, the transition of $F^{1/n}$ from the linear regime to the plateau exhibits a sharp change at the critical value $p_{\rm c}\simeq 0.5$, indicating the occurrence of a phase transition. The internal energy shows a discontinuous jump from zero to a finite value $E/n\simeq 0.5$ at $p_{\rm c}$, which reflects a sudden proliferation of Pauli errors. The discontinuity of the internal energy, which corresponds to the latent heat for the classical spins, indicates that the phase transition is of {\it first-order}. The specific heat displays a pronounced peak at the same point. As the number of qubits increases, the jump in internal energy becomes more abrupt and the specific heat peak becomes sharper and higher, unlike the case of the 2D cluster state, where the peak remains broad and size-independent (see Figs.~\ref{fig:2Dgraphstate}(e) and (f)).

In the presence of a phase transition, the pure state term $(1-p)^n$ in Eq.~(\ref{eq:fidelity}) dominates for $p$ smaller than $p_{\rm c}$, while for $p\ge p_c$ the contribution from Pauli noise, which is the second term in Eq.~(\ref{eq:fidelity}), abruptly becomes significant, driving the system toward the maximally mixed state. The phase transition arises from this sudden onset of noise effects.

Whether a phase transition occurs depends critically on the degree of the graph state. Figures~\ref{fig:2D_spheat} (a) and (b) show the specific heat for 2D $d$-regular graph states with $3\le d\le 8$. 
As $d$ increases, the peak of the specific heat becomes sharper and more pronounced. 
For graph states with $d\le 5$, the peak remains broad and low, indicating the absence of a phase transition. 
In contrast, for $d\ge 6$, the specific heat grows and sharpens with increasing system size, clearly signaling the onset of a phase transition. These results suggest that phase transitions occur when the degree exceeds 5, while no transition is observed for degrees below 6.

When a phase transition occurs, since the fidelity is dominated by the pure state term $F=(1-p)^n$ for $p< p_{\rm c}$, it decreases more rapidly than in cases without a phase transition.
Consequently, graph states with lower connectivity ($d\le 5$), which exhibit a smooth crossover between the pure and maximally mixed states, are more robust against noise than highly connected ones ($d\ge 6$).

Interestingly, when the degree becomes excessively large, the phase transition disappears. 
For example, the fidelity for an $n$-qubit fully connected graph state, given by \cite{TTYYT23} (see also Sec.~\ref{subsec:constraint-percolation})
\begin{align}
F=\frac{1}{2}\left[\left(1-\frac{2}{3}p\right)^n+\left(\frac{2}{3}p\right)^n+\left(1-\frac{4}{3}p\right)^n\right]
\label{fulgraph_fidexact}
\end{align}
does not exhibit any singularity in the limit $n\to\infty$.
Determining the maximum degree at which a phase transition can still occur remains a question for future work.

\subsection{3D graph states}

Figures~\ref{fig:3Dgraphstate} (a)–(f) present the results for the 3D 5-regular graph state and the 3D cluster state. 
While the 2D 5-regular graph state does not exhibit a phase transition, the 3D 5-regular graph state shows clear signs of a phase transition, as evidenced by the growing peak in the specific heat with increasing system size. 
This indicates that the occurrence of a phase transition depends not only on the degree but also on the dimensionality of the graph state. 
The results for the 3D cluster state, shown in Figs.~\ref{fig:3Dgraphstate} (d)–(f), further confirm the presence of a phase transition and reinforce the trend that phase transitions become more likely as degree increases.
These results indicate that the dimensionality of the graph state significantly affects its robustness. That is, higher-dimensional graph states are more fragile to noise than low-dimensional ones.

\subsection{Entanglement-breaking nature of depolarizing noise}

In Figs.~\ref{fig:2D_spheat} (b), \ref{fig:3Dgraphstate} (c), and \ref{fig:3Dgraphstate} (f), a phase transition is observed at the critical probability $p_{\rm c}\simeq 0.5$, which is independent of both the degree of the graph and the spatial dimensionality.
This universality of the critical point does not originate from special properties of graph states, but rather may arise from the entanglement-breaking nature of IID depolarizing noise. 

It is known that the single-qubit depolarizing channel becomes entanglement-breaking for $p\ge 1/2$ \cite{MZ10}, meaning that it destroys entanglement between the qubit and any external reference system, and in particular maps any two-qubit state to a separable state in this parameter regime. Since entanglement-breaking channels can always be written in the Holevo (measure-and-prepare) form \cite{HSR03}, the tensor product of entanglement-breaking channels is also entanglement breaking, as it can be written in the same form. Consequently, the IID depolarizing channel acting on multiple qubits is entanglement breaking for $p\ge 1/2$, and we therefore conjecture that the fidelity undergoes a phase transition at this point, irrespective of the underlying graph geometry.

\subsection{Mapping to a constraint-percolation problem}
\label{subsec:constraint-percolation}

To understand the qualitative difference between two-dimensional regular graph states with degrees $d=5$ and $d=6$ as well as the unusual behavior of fully connected graph states, we make use of the fact that the fidelity of graph states subject to IID depolarizing noise can be mapped exactly onto the partition function of a constraint-percolation problem (see Appendix~B for derivation),
\begin{align}
&F=\frac{1}{2^n}\sum_{\{s\}}\prod_{i=1}^n[(1-p')+p'D_i]\label{eq:fidelity_percolation}\\
&=\frac{1}{2^n}\sum_{l=0}^n(p')^l(1-p')^{n-l}\sum_{i_1,i_2,\cdots,i_l}\sum_{\{s\}}\left(\prod_{k=1}^l D_{i_k}\right),
\label{eq:percolation2}
\end{align}
where $\sum_{\{s\}}=\prod_{j=1}^n\sum_{s_j=\pm1}$, $p'=4p/3$, $A_i=\prod_{j:(i,j)\in E} s_j$, and $D_i=\delta_{s_i,1}\delta_{A_i,1}$. 

Equation~(\ref{eq:percolation2}) has a structure analogous to the partition function of the Fortuin–Kasteleyn random-cluster model \cite{W82}. The fidelity in Eq.~(\ref{eq:percolation2}) can be viewed as the sum of probabilities over all spin configurations. The probability of a configuration is determined by the rule that the spin at the $i$-th vertex contributes a weight $p'$ if $D_i=1$ ($s_i=1$ and $A_i=1$), and a weight $1-p'$ otherwise. In the constraint-percolation model picture, we regard a vertex as ``occupied" if $D_i=1$, and as ``empty" otherwise. Thus, each vertex is occupied with probability $p'$ only when it is connected to an even number of spin $-1$s, and is otherwise empty with probability $1-p'$.
As described in Appendix B, Eq.~(\ref{eq:percolation2}) can be interpreted as describing the percolation of identity operators within products of stabilizers.

From this perspective, the original problem of evaluating the fidelity of graph states under IID depolarizing noise reduces to analyzing the statistical mechanics of constrained clusters of occupied vertices on the underlying graph.

We conjecture the following explanation for the qualitative difference between two-dimensional graph states with degrees $d=5$ and $d=6$ in terms of constraint percolation: We begin by discussing the case $2\le d\le4$. In this case, the constraints $D_i=1$ allow only a family of geometrically restricted compact clusters of occupied vertices, so that increasing the noise strength does not cause a sudden decrease of compatible spin configurations. Instead, the dominant contributions to the fidelity change gradually with $p$, and the transition between the low-noise and high-noise regimes appears as a smooth crossover rather than a phase transition. 
For $d=6$, however, the local constraints $D_i=1$ admit compact clusters and macroscopic clusters that spread the entire system and appear in various sizes. These macroscopic clusters of occupied vertices yield a threshold of the noise strength $p_{\rm c}$ such that, once $p$ exceeds it, macroscopic clusters start to emerge, drastically reducing the number of spin configurations compatible with the constraints. As a result, the fidelity rapidly approaches the maximally mixed value $F\simeq 2^{-n}$, leading to a sharp phase transition.
Meanwhile, for graph states with $d=5$, both compact clusters and macroscopic clusters are allowed by the local constraints. However, since the macroscopic clusters are restricted to limited sizes due to geometric constraints and the smallest macroscopic cluster is larger than that for $d=6$, near $p=1/2$ only compact clusters contribute to the fidelity significantly, and the fidelity changes continuously. Macroscopic clusters appear only at higher noise strengths, and therefore a sharp phase transition does not occur.

Fully connected graph states exhibit qualitatively different behavior compared to graph states with low connectivity. In this case, the local constraint $D_i=1$ does not restrict the occupation of vertices in a local manner, because every vertex is connected to all others. The condition $A_i=1$ at any occupied vertex reduces to a single global parity condition on the entire configuration, $P=\prod_i s_i=\pm 1$. In the case $P=+1$, where the number of spin $-1$ is even, every vertex with $s_i=+1$ satisfies the constraint $D_i=1$, because it is connected to an even number of $-1$ spins and therefore $A_i=1$. By contrast, it is not satisfied at vertices with $s_i =-1$. 
From Eq.~(\ref{eq:fidelity_percolation}), the contribution to the fidelity from configurations with $P=1$ is given as
\begin{align}
F_{P=1}&=\frac{1}{2^n}\sum_{\substack{k:{\rm even} \\ 0\le k\le n}}\left(
\begin{array}{c}
n\\
k
\end{array}
\right)1^{n-k}(1-p')^k\\
&=\frac{1}{2}\left[\left(1-\frac{p'}{2}\right)^n+\left(\frac{p'}{2}\right)^n\right],
\end{align}
where we have used the identity given in Eq.~(29) of Ref.~\cite{TTYYT23}.
On the other hand, in the case $P=-1$, where the number of $-1$ spins is odd, the conditions $s_i=1$ and $A_i=1$ cannot be satisfied simultaneously at any vertex. The fidelity contribution from these configurations is therefore
\begin{equation}
F_{P=-1}=\frac{1}{2^n}(1-p')^n\sum_{\substack{k:{\rm odd} \\ 1\le k\le n}}
\left(
\begin{array}{c}
n\\
k
\end{array}
\right)=\frac{1}{2}(1-p')^n.
\end{equation}
We thus obtain the fidelity of fully connected graph states as 
\begin{align}
F&=F_{P=+1}+F_{P=-1}\\
&=\frac{1}{2}\left[\left(1-\frac{p'}{2}\right)^n+\left(\frac{p'}{2}\right)^n+\left(1-p'\right)^n\right].
\label{eq:fulgraph_fid_percolation}
\end{align}
Substituting $p'=4p/3$, Eq.~(\ref{eq:fulgraph_fid_percolation}) indeed becomes Eq.~(\ref{fulgraph_fidexact}).

The above derivation of the fidelity of fully connected graph states shows that local constraints arising from different vertices do not accumulate as independent constraints, but instead become globally redundant as the cluster of occupied sites grows. As a result, even large clusters do not significantly reduce the number of spin configurations of compatible with the constraints. 
Consequently, although the underlying graph is maximally connected, increasing the noise does not lead to a sudden decrease in the number of spin configurations or to a sharp transition in the fidelity. Instead, the crossover from the pure-state regime to the noise-dominated regime remains smooth, demonstrating that extreme connectivity suppresses critical behavior by making the constraints globally redundant.

\medskip
\section{Conclusion and discussion
\label{VI}}

In this work, we demonstrated that the fidelity of any graph state under IID Pauli noise can be mapped to the partition function of a classical spin system, which enables efficient evaluation of fidelity via statistical mechanical methods.

We found that fidelity undergoes a sharp phase transition driven by the abrupt onset of Pauli noise effects at the critical value of probability $p_{\rm c}\simeq 0.5$ for graph states on regular graphs with sufficiently high degree and dimensionality. Specifically, phase transitions appear for $d\ge 6$ in 2D and $d\ge 5$ in 3D. For 1D and lower-degree graphs in 2D, the fidelity exhibits a smooth crossover, indicating the absence of a phase transition.

We further found that robustness against noise is determined by the graph structure. 
Graph states with lower connectivity and dimensionality, which display a smooth crossover, tend to be more robust, while those with higher connectivity or dimensionality are more fragile. However, in the case of extreme connectivity—as in fully connected graphs—critical behavior is suppressed and robustness is restored.

Finally, we have shown that the fidelity can be written as the partition function of a constraint-percolation problem. Based on this picture, we have provided possible explanations for why, in two dimensions, the regular graph state with $d=6$ shows a phase transition while that with $d=5$ exhibits a smooth crossover, and why fully connected graph states display suppressed critical behavior despite their high connectivity.

Several open questions remain. First, while this study focused on depolarizing noise, it would be valuable to extend the analysis to other types of IID noise and correlated noise models. 
Second, the maximum degree of a regular graph state that can exhibit a fidelity phase transition remains to be determined. 
Third, it would be important to investigate whether the mapping of fidelity to the partition function of a classical spin model can be extended to other classes of stabilizer states.
Fourth, an interesting open problem is to develop a more rigorous understanding of the relation between the universality of the critical probability and the entanglement-breaking nature of IID depolarizing noise. It is also important to clarify the distinction between regular graph states with $d=5$ and $d=6$ and the unusual behavior of fully connected graph states, particularly how the geometry of constraint clusters determines the presence or absence of a phase transition.
Finally, exploring the connection between noise robustness and computational ability as a resource state of MBQC is an interesting direction for future research.

\medskip

\section*{APPENDIX A: Mean-field approximation}
\label{A}

We perform a MF analysis to the classical spin system obtained by mapping the fidelity of graph states. We consider graph states under depolarizing noise that correspond to the Hamiltonian Eq.~(\ref{eq:depolarizingH}). 

Classical spin $s_i$ can be written as $s_i=\langle s\rangle+\delta s_i$, where $\langle s\rangle$ denotes the average of $s_i$ and $\delta s_i$ denotes fluctuation from the average, which is assumed small.
Within the first order of fluctuation,  a product of spins can be written as
\begin{align}
s_{i_1}s_{i_2}\cdots s_{i_l}=\langle s\rangle^l+\langle s\rangle^{l-1} \sum_{j=1}^l\delta s_{i_j}\notag\\
=(1-l)\langle s\rangle^l+\langle s\rangle^{l-1}\sum_{j=1}^ls_{i_j}.
\end{align}
Using this formula, the MF Hamiltonian can be written as
\begin{equation}
\mathcal H_{\rm MF}=B(\langle s\rangle)\sum_{i=1}^n s_i+D(\langle s\rangle),
\end{equation}
where the form of the functions $B(\langle s\rangle)$ and $D(\langle s\rangle)$ depend on the graph. 
For instance, their explicit forms for $d$-dimensional cluster states ($d=1,2,3$) are given as
\begin{align}
B(\langle s\rangle)=\frac{1}{4}-\frac{k}{4}\langle s\rangle^{k-1}+\frac{k+1}{4}\langle s\rangle^{k}, 
\label{eq:meanfieldB}\\
D(\langle s\rangle)=\frac{3n}{4}+\frac{n(k-1)}{4}\langle s\rangle^{k}-\frac{nk}{4}\langle s\rangle^{k+1},
\label{eq:meanfieldD}
\end{align}
where $k\equiv 2d$ is the degree of $d$-dimensional cluster state.

The partition function of the mean-field Hamiltonian $\mathcal Z_{\rm MF}$ is given as
\begin{align}
\mathcal Z_{\rm MF}&=\sum_{s_1=\pm 1}\sum_{s_2=\pm1}\dots\sum_{s_n=\pm 1} e^{-\beta({\mathcal H}_{\rm MF}+c)}\notag\\
&=e^{-\beta(D+c)}2^n\cosh^n(\beta B).
\end{align}
Thus, the self-consistent equation for $\langle s\rangle$ is given as
\begin{align}
\langle s\rangle&=\sum_{s_1=\pm 1}\sum_{s_2=\pm1}\dots\sum_{s_n=\pm 1}s_i
\frac{e^{-\beta({\mathcal H}_{\rm MF}+c)}}
{{\mathcal Z}_{\rm MF}}\notag\\
&=-\tanh(\beta B). 
\end{align}
Solving the above self-consistent equation, magnetization $\langle s\rangle$ can be determined. 
If multiple solutions are found, the stable solution has minimum free energy. The mean-field free energy is given as
\begin{equation}
    {\mathcal F}_{\rm MF}=D+c-\frac{n}{\beta}\ln\left[2\cosh(\beta B) \right].
\end{equation}

\section*{APPENDIX B: Mapping to a constraint-percolation problem}
\label{B}

We show that the fidelity of a graph state subject to IID depolarizing noise can be mapped exactly to the partition function of a constraint-percolation problem.
We start from the expression for the fidelity of a graph state under IID depolarizing noise (see Eq.~(23) in Ref.~\cite{TTYYT23})
\begin{align}
F=\frac{1}{2^n}\sum_{\ell\in\{0,1\}^n}\left(1-p'\right)^{m(\ell)},
\label{eq:F_tanizawa}
\end{align}
where $p'=4p/3$, and $m(\ell)$ denotes the number of non-identity Pauli operators appearing in the stabilizer specified by the bit string $\ell$.
The summation over $\ell\in\{0,1\}^n$ can be interpreted as a summation over classical spin configurations,
\begin{equation}
%\sum_{\ell\in\{0,1\}^n}\leftrightarrow
\sum_{\{s\}}=\prod_{j=1}^n\sum_{s_j=\pm 1},
\end{equation}
where, for a given $\ell$, the value of $m(\ell)$ coincides with that of the classical Hamiltonian $\mathcal H(\{s\})$ defined in Eq.~(\ref{eq:depolarizingH}) for the corresponding spin configuration $\{s\}$. For later convenience, we perform the transformation $s_i\to -s_i$ in Eq.~(\ref{eq:depolarizingH}) throughout this Appendix. 
With this convention, Eq.~(\ref{eq:F_tanizawa}) can be written as
\begin{align}
F&=\frac{1}{2^n}\sum_{\{s\}}e^{K\mathcal H(\{s\})},\label{F_tanizawa2}\\
H(\{s\})&=\sum_{i=1}^n\left[\frac{1}{2}(1-s_i)+\frac{1}{4}(1+s_i)(1-A_i)\right],
\end{align}
where $K\equiv\ln (1-p')$ and $A_i\equiv\prod_{j:(i,j)\in E} s_j$. 
Using the identities 
\begin{align}
e^{K(1-s_i)/2}&=1+(e^K-1)\delta_{s_i,-1}\notag\\&=1-p'\delta_{s_i,-1},\\
e^{K(1+s_i)(1-A_i)/4}&=1+(e^K-1)\delta_{s_i,1}\delta_{A_i,-1}\notag\\
&=1-p'\delta_{s_i,1}\delta_{A_i,-1},
\end{align}
the weight in Eq.~(\ref{F_tanizawa2}) factorizes as
\begin{align}
e^{K\mathcal H(\{s\})}&=\prod_{i=1}^n(1-p'\delta_{s_i,-1})(1-p'\delta_{s_i,1}\delta_{A_i,-1})\\
&=\prod_{i=1}^n\left[1-p'(\delta_{s_i,-1}+\delta_{s_i,1}\delta_{A_i,-1})\right]\\
&=\prod_{i=1}^n[(1-p')+p'\delta_{s_i,1}\delta_{A_i,1}].
\label{eq:weight}
\end{align}
The fidelity therefore reduces to Eq.~(\ref{eq:fidelity_percolation}) as
\begin{align}
F&=\frac{1}{2^n}\sum_{\{s\}}\prod_{i=1}^n[(1-p')+p'D_i]\\
&=\left(\frac{1-p'}{2}\right)^n\sum_{\{s\}}\prod_i(1+xD_i),
\label{eq:percolation}
\end{align}
where we have defined $D_i\equiv \delta_{s_i,1}\delta_{A_i,1}$ and $x\equiv p'/(1-p')$. 
Expanding the product in Eq.~(\ref{eq:percolation}), we obtain Eq.~(\ref{eq:percolation2}) as
\begin{align}
F=\frac{1}{2^n}\sum_{l=0}^n(p')^l(1-p')^{n-l}\sum_{i_1,i_2,\cdots,i_l}\sum_{\{s\}}\left(\prod_{k=1}^l D_{i_k}\right).
%\label{eq:percolation2}
\end{align}
Since $s_i=+1$ represents the absence of the stabilizer generator $g_i$ at the $i$-th qubit, and $A_i=\prod_{j:(i,j)\in E} s_j=1$ ensures that an even number of neighboring generators act on that qubit, Eq.~(\ref{eq:percolation2}) can be interpreted as describing the percolation of identity operators within stabilizers.

To check the correctness of Eq.~(\ref{eq:percolation2}), we evaluate the fidelity in the two limiting regimes $p'\ll 1$ and $1-p'\ll 1$. In the weak noise limit $p'\ll 1$, the expansion in Eq.~(\ref{eq:percolation2}) is dominated by the $l=0$ and $l=1$ terms, both of which give rise to first order terms in $p'$. Expanding these two terms to first order in $p'$, we obtain
\begin{align}
F_0&=(1-p')^n=1-np'+O(p'^2),\\
F_1&=\frac{1}{2^n}np'(1-p')^{n-1}\sum_{\{s\}}D_i\notag\\
&=\frac{1}{4}np'(1-p')^{n-1}=\frac{n}{4}p'+O(p'^2),
\label{eq:F1}
\end{align}
where $F_0$ and $F_1$ correspond to the contributions from the $l=0$ and $l=1$ terms, respectively. In deriving Eq.~(\ref{eq:F1}), we have used 
\begin{align}
\sum_{\{s\}}D_i=\sum_{\{s\}}\delta_{s_i,1}\delta_{A_i,1}=2^{n-2}.
\end{align}
Combining these contributions, the fidelity in the weak noise limit is given by
\begin{equation}
F=F_0+F_1=1-\frac{3}{4}np'\simeq \left(1-\frac{3}{4}p'\right)^{n}=(1-p)^n,
\end{equation}
as expected.
In the opposite limit, the dominant contribution comes from the term with $l=n$, corresponding to the configuration in which all spins are fixed to $+1$ and all constraints are satisfied. In this case, since the number of compatible spin configurations to constraints is $O(1)$, the fidelity approaches the maximally mixed value $F\simeq 1/2^n$.

\section*{ACKNOWLEDGMENTS}
ST is grateful to K. Fujii and D. Kagamihara for their insightful comments. ST is supported by JSPS KAKENHI Grant Number 25K07158.
SY is supported by Grant-in-Aid for JSPS Fellows (Grant No. JP22J22306) and Grant-in-Aid for Young Scientists (Start-up) No. 25K23355.
RY is supported by JSPS KAKENHI Grant Numbers~20H01838 and 25K07156 and the WPI program ``Sustainability with Knotted Chiral Meta Matter (SKCM$^2$)'' at Hiroshima University. 
YT is partially supported by the MEXT Quantum Leap Flagship Program (MEXT Q-LEAP) Grant Number JPMXS0120319794 and JST [Moonshot R\&D -- MILLENNIA Program] Grant Number JPMJMS2061.

%\medskip

\end{document}